\documentclass[onecolumn,12pt]{article}
\usepackage{a4wide}
\usepackage{multicol}
\usepackage{amssymb}
\usepackage{graphicx}
\usepackage{subfigure}

\begin{document}

\newcommand{\be}{\begin{equation}}\newcommand{\ee}{\end{equation}}
\newcommand{\bea}{\begin{eqnarray}} \newcommand{\eea}{\end{eqnarray}}

\rightline{ICCUB-11-143}

\bigskip

\begin{center}

{\Large\bf  Fuzzy spheres  at finite temperature\\
 supported by Wilson lines
}

\bigskip
\bigskip

{ \large Diego Rodriguez-Gomez \footnotemark[1] \footnotemark[2] and Jorge G. Russo \footnotemark[3] \footnotemark[4]}
\bigskip

{\it
1) Department of Physics, \\
Technion, Haifa, 3200, Israel\\
\smallskip
2) Department of Mathematics and Physics, \\
University of Haifa at Oranim, Tivon, 36006, Israel\\
\smallskip
3) Instituci\'o Catalana de Recerca i Estudis Avan\c cats (ICREA)\\
Pg. Lluis Companys, 23, 08010 Barcelona, Spain\\
\smallskip
4) Institute of Cosmos Sciences and Estructura i Constituents de la Materia\\
Facultat de F{\'\i}sica, Universitat de Barcelona\\
Av. Diagonal 647,  08028 Barcelona, Spain\\
}
\bigskip
\bigskip

\end{center}
\bigskip

\begin{abstract}

 We study the fate of the fuzzy sphere vacuum solutions in the  $SU(N)\times SU(N+M)$ Klebanov-Strassler theory at finite temperature and at weak coupling. We find that thermal effects push the $S^2$ radius --a modulus at $T=0$-- towards vanishing size. We then show that the fuzzy sphere can be stabilized by  adding a temporal Wilson line for a background $U(1)$ gauge field, which can be associated with either the baryonic or the R-symmetry. Thus, upon deforming the KS theory with a background gauge field for a global symmetry, we find a phase of broken baryonic symmetry at weak coupling surviving at finite temperature.

\end{abstract}

\clearpage

\section{Introduction}

Due to its comparative simplicity and yet generality, the gauge/gravity duality arising from D3 branes probing a conifold singularity  has grown to become one of the most extensively studied scenarios. When no fractional branes are present the system flows to a non-trivial IR strongly coupled fixed point, whose gravity description is in terms of classical IIB supergravity on $AdS_5\times T^{1,1}$. The field theory side is the  Klebanov-Witten (KW) quiver field theory \cite{Klebanov:1998hh} with gauge group $SU(N)\times SU(N)$. On the other hand, when $M$ fractional branes are present, the gauge group is $SU(N)\times SU(N+M)$ and it is commonly known as the Klebanov-Strassler (KS) theory \cite{Klebanov:2000hb}. In this case the system flows to a confining theory and a rich dynamic emerges. In turn, the gravity dual (on the baryonic branch) to the $SU(Mk)\times SU(M(k+1))$ is in terms of a deformed conifold geometry with certain fluxes  \cite{Klebanov:2000hb, Butti:2004pk}.

In \cite{Dymarsky:2005xt} a comprehensive study of the various branches of the moduli space of the KS theory was performed. In particular, an interesting solution to the D and F flatness conditions along the baryonic branch was found. It was later understood in \cite{Maldacena:2009mw} that it represented a fuzzy 2-sphere, whose role in elucidating the nature of the geometric transition from a purely gravitational perspective was found crucial.

The radius of this $S^2$ has an interesting interpretation as the scalar component in the supermultiplet of the current associated with the baryon symmetry of the theory. In fact, this operator can be thought of as the Fayet-Iliopoulos parameter of the abelian part of the gauge symmetry if we considered the $U(N)\times U(N+M)$ theory. However, in the $SU(N)\times SU(N+M)$  case, it remains as a modulus. In this note we are interested in studying  thermal effects on such configuration. We will see that they reveal interesting aspects of the dynamics.
First, we find that finite temperature pushes the fuzzy sphere radius towards zero. We then show that
by turning on a temporal Wilson line for a background $U(1)$ gauge field, there is a region of parameters where the fuzzy sphere is stabilized, leading to a phase of broken baryonic symmetry surviving at high temperatures. The mechanism is very similar to the Hosotani mechanism \cite{Hosotani:1983xw} to generate dynamical gauge symmetry breaking in gauge theories by adding Wilson lines in extra dimensions.
The key observation is that turning on this temporal Wilson line $A$ is equivalent to having an imaginary chemical potential. Thus, in the partition function the terms for \textit{e.g.} bosons $\log\big|1-e^{-\beta\, (E-\mu)}\big|$ will turn into $\log \big|1-e^{-\beta\, (E-i\, q\,A)}\big|$. 
The stabilization mechanism is then simply exhibited for $\beta\,q\,A=\pi $, when the Bolzmann factor will acquire an extra sign, thus making formally the bosons to contribute as fermions and vice versa. However, the overall global sign of the fermionic and bosonic contributions will not be affected by this, and so the net effect is  a thermal contribution with a sign opposite to the $A=0$ case, which in a regime of parameters leads to an absolute minimum at some $v\neq 0$, rendering   the $v=0$
vacuum metastable.  

 An alternative way of looking at this phenomenon is by noticing that the background gauge field is a way to implement twisted boundary conditions in fermions and bosons by means of a physical configuration.\footnote{Twisted boundary conditions have been first used in \cite{Scherk:1978ta, Scherk:1979zr} as a mechanism for (super) symmetry breaking.} From this perspective, the particular value $\beta\ q\,A=\pi$ makes fermions periodic and bosons antiperiodic, hence triggering the  instability discussed above. It should be noted, however, that the value $\beta\ q\,A=\pi$ is by no means special, in fact the effect exists in a number of intervals for the parameter $a\equiv \beta\, q\,A$, whose extensions  depend on the coupling constant.  

The stabilization of the fuzzy sphere proposed in this note might have interesting cosmological applications. One can imagine a cosmological scenario where the fuzzy sphere traverses various epochs, which bears some similarities with the scenario of \cite{DeWolfe:2004qx} (see also \cite{Ashoorioon:2009wa}), in the present case however as temperature is lowered.

The organization of this note is as follows. In section \ref{Backgr} we will briefly review the fuzzy $S^2$ vacuum of the KS theory. In section \ref{instabilities} we show that at finite temperature the fuzzy sphere shrinks to zero radius, but that it can be  stabilized at a non-zero radius by adding a temporal Wilson line. This is done by deforming the $SU(N)\times SU(N+M)$ theory by the addition of a background gauge field for an  otherwise global symmetry.
We conclude in section \ref{conclusions} with some future directions and possible applications of our mechanism. 

\section{The fuzzy sphere in KS}\label{Backgr}

Following \cite{Maldacena:2009mw}, we focus on the $SU(M\,k)\times SU(M\,(k+1))$ theory, whose action in $\mathcal{N}=1$ superspace is

\begin{equation}
S=\int {\rm Tr}\Big\{\tau_1\,\mathcal{W}_1\,^2+\tau_2\,\mathcal{W}_2\,^2+\sum A_i^{\dagger}\, e^{-V_1}\, A_i\,e^{V_2} +\sum B_i^{\dagger}\, e^{-V_2}\, B_i\,e^{V_1}+W\,\Big\}\, ,
\label{uno}
\end{equation}
where $\tau_1,\ \tau_2$ are the holomorphic couplings of the vector multiplets $V_1,\, V_2$ of the $SU(N)\times SU(N+1)$ gauge symmetry, and $A_i,\,B_i$, $i=1,2$, are the bifundamental fields transforming respectively as $(\Box,\,\bar{\Box})$ and $(\bar{\Box},\,\Box)$ . In turn, the superpotential is

\begin{equation}
W=\frac{\lambda}{2}\,A_i\,B_m\,A_j\,B_n\,\epsilon^{ij}\,\epsilon^{mn}\ .
\end{equation}

The theory has an $SU(2)_1\times SU(2)_2\times U(1)_B\times U(1)_R$ global symmetry group. The corresponding charges are
as follows 

\begin{equation}
\begin{array}{c | c c c} 
& A_i & & B_i \\ \hline
U(1)_R &  \frac{1}{2}&  &\frac{1}{2}\\
U(1)_B & 1& & -1  \\
SU(2)_1 &&& \Box  \\
SU(2)_2 & && \Box
\end{array}
\end{equation}
We note that the R-charge labelling the chiral multiplet is that of the scalar component. Because of supersymmetry, the R-charge of the fermionic component is $R[{\rm scalar}]-1$. Thus, the fermions have $R$-charge $-\frac{1}{2}$. On the other hand, the baryonic symmetry, as it commutes with the supercharges, assigns the same charge to bosons and fermions in the multiplet.

The fuzzy sphere vacuum solution reads as follows \cite{Maldacena:2009mw}

\begin{equation}
\label{vacuum}
A_1= v\,\left(\begin{array} { c c c c c} \sqrt{N} & 0 & \cdots & 0 & 0 \\ 0 & \sqrt{N-1} & \cdots & 0 & 0 \\ 0 & 0 & \cdots & 1 & 0 \end{array} \right)\, , \qquad 
A_2= v\,\left(\begin{array} { c c c c c} 0& 1 & \cdots & 0 & 0 \\ 0 & 0 & \sqrt{2} & \cdots  & 0 \\ 0 & 0 & \cdots & 0 & \sqrt{N} \end{array} \right)\,,
\end{equation}
where each entry is an $M\times M$ identity matrix and $B_i=0$. 
For our purposes, we can set for simplicity $M=1$, as the $M$-dependence will be  an overall factor 
(or alternatively just choose $M=1$).  We also rename $k=N$. Because the $B_i$ fields are set to zero, the F-term equations are trivially satisfied. Furthermore, one can see that

\begin{equation}
\label{D-terms}
\sum_i\, A_i\,A_i^{\dagger} = (N+1)\,v^2\,\mathbf{1}_N \ , \qquad \sum_i\,A^{\dagger}_i\,A_i=N\,v^2\,\mathbf{1}_{N+1}\ .
\end{equation}
Thus, it is immediate to see that the configuration is also D-flat for any $v$, which is therefore a modulus. 
In this fuzzy sphere vacuum the baryonic symmetry is broken; a baryonic operator of the rough form $\mathfrak{B}[A]=A_1^{\frac{N\,(N+1)}{2}}\,A_2^{\frac{N\,(N+1)}{2}}$ gets a vacuum expectation value (we refer to \cite{Dymarsky:2005xt} for further details).

In the following we will be interested in studying the ultraviolet regime of the field theory; that is, energies well above the strong coupling scale $\Lambda_i$ of either node. Since the gauge fields are asymptotically free we can basically neglect the gauge sector, thus keeping the Wess-Zumino model for the fields $A_i,\,B_i$ with superpotential $W$. The configuration (\ref{vacuum}) clearly represents  a supersymmetric vacuum of this simplified model. 

\section{Perturbative instabilities at finite temperature due to Wilson lines}\label{instabilities}

We are now interested in coupling the KS theory to a background gauge field for an abelian global symmetry. This background gauge field can be thought of as a deformation of the original $SU(M\,k)\times SU(M\,(k+1))$ gauge theory. We note that this type of deformation is well known in a 3-dimensional context, as it encodes real masses for chiral multiplets (see \textit{e.g.} \cite{Aharony:1997bx}). Very recently its 4d (that is, our case) realization as well as  the relation with the aforementioned 3d real mass have been considered in \cite{Festuccia:2011ws}. Deformations whereby one adds a background gauge field for an  abelian symmetry have also been very recently considered in \cite{Maeda:2010br}, also from a gravitational perspective.

The background abelian gauge field is added by changing the derivatives in the kinetic terms of the Wess-Zumino model as

\begin{equation}
\partial_{\mu}\rightarrow \partial_{\mu}-i\,q\,A_{\mu}\, ,
\end{equation}
where $|q|=1,\,\frac{1}{2}$ depending on whether we use the baryonic or R-symmetry. 
We will assume our background gauge field to have only $A_0$ non-vanishing, which we will be further assumed to be constant. 
At finite temperature, where  euclidean time is compactified into a circle of radius $\beta =T^{-1}$, a constant potential $A_0$ will have an important physical effect:  it will lead to twisted boundary conditions for fermions and bosons in the thermal circle, giving
rise to a rich vacuum dynamics.

We now turn to the computation of the one-loop effective potential. Following the classical reference \cite{Dolan:1973qd} we need to expand our bosonic and fermionic fields around the vacuum configuration  (\ref{vacuum}) and keep the terms which are quadratic in the fluctuations. 
Linear terms are  absent in the present case (in general cases, they can be re-absorbed by a wave function renormalization of the classical field). 
The one-loop effective potential will thus have a classical contribution plus the one-loop contribution coming from the fluctuations. This one-loop piece splits into a  Coleman-Weinberg (which is independent of $A_0$) and a thermal contribution (which depends on $A_0$).  Because the configuration is supersymmetric, there is an equal number of complex scalar fluctuations as of Majorana fermionic fluctuations with a given  mass. Therefore, the Coleman-Weinberg quantum contribution cancels out among fermions and bosons, and the effective potential to one-loop order is just given by the sum of the classical plus the one-loop thermal contribution. This contribution is straightforward to work out once the spectrum of fluctuations is known. To that matter, due to supersymmetry, it is enough to keep track of the bosonic piece of the fluctuations spectrum. It is not hard to check that for the $SU(N)\times SU(N+1)$ theory, upon diagonalizing the mass-matrix, we  have

\begin{itemize}
\item $4\,n$ complex scalars of $m^2=n^2\,\lambda^2\,v^4$ for $n=1,\cdots , N$.
\item $2\,N\,(N+1)$ massless complex scalar fluctuations.
\end{itemize}
The contribution of each fluctuation multiplet of mass $m$ to the one-loop thermal potential is

\begin{equation}
\label{DeltaV}
\Delta V_{\rm eff}=\frac{1}{\beta}\,\int \frac{d^3\vec{p}}{(2\pi)^3}\,\log\Big(\left|\frac{1-e^{-\beta\,(E+i\,q\,A)}}{1+e^{-\beta\,(E+i\,q\,A)}}\right|^2\Big)\ ,
\end{equation}
where $E^2=\sqrt{\vec{p}\,^2+m^2}$ and we have set $A_0\equiv  A$. In this expression the numerator in the argument of the log stands for the contribution of the boson in the multiplet, while the denominator represents the fermion. Furthermore the absolute value is due to the fact that for both boson and fermion we have to sum over particle and antiparticle, and thus add the two contributions with opposite signs of $q$. Here $A$ can be associated with either the baryon or R-symmetry $U(1)$.
Because  both R-charges and baryon charges are equal in absolute value for fermions and bosons in the multiplet, the  contribution 
(\ref{DeltaV}) has the same form in both cases, with $q=1$ and $q=1/2 $ for baryon and R-symmetry respectively.

Since $|e^{-\beta\,(E+i\,q\,A)}|<1$, we can expand the logarithm as

\begin{eqnarray}
 \log\Big(\left|\frac{1-e^{-\beta\,(E+i\,q\,A)}}{1+e^{-\beta\,(E+i\,q\,A)}}\right|^2\Big)= -4\sum_{k=0}^{\infty}\, \frac{\cos(q\,A\,\beta\,(2k+1))}{(2k+1)}\,e^{-\beta\, (2k+1)\,E}\ .
\end{eqnarray}
Performing the momentum integration, we find that a massless multiplet contributes

\begin{equation}
\Delta V_{\rm eff}^{(m=0)}=-\frac{4}{\pi^2\,\beta^4}\,\sum_{k=0}^{\infty}\,\frac{\cos(q\,A\,\beta\,(2k+1))}{(2k+1)^4}\ ,
\label{masacero}
\end{equation}
whereas a massive multiplet contributes

\begin{equation}
\Delta V_{\rm eff}^{(m)}=-\frac{2\,m^2}{\pi^2\,\beta^2}\,\sum_{k=0}^{\infty}\,\frac{\cos(q\,A\,\beta\,(2k+1))}{(2k+1)^2}\,K_2(m\,\beta\,(2k+1))\ .
\end{equation}
In addition, there is a classical contribution given by

\begin{equation}
V_{\rm eff}^{cl}=q^2\,A^2\,v^2\,N\,(N+1)\ .
\end{equation}
Thus, taking into account the complete fluctuation spectrum, we finally find the following expression for the one-loop effective potential

\be
V_{\rm eff}^{\rm total}=V_{\rm eff}+2\,N\,(N+1) \Delta V_{\rm eff}^{(m=0)}\ ,
\ee
with

\begin{eqnarray}
\label{V}
V_{\rm eff}=\frac{a^2\,v^2\,N\,(N+1)}{\beta^2}
-\frac{8\,\lambda_0^2\,v^4}{\pi^2}\sum_{k=0}^{\infty}\,\frac{\cos(a\,(2k+1))}{(2k+1)^2}\, \sum_{n=1}^{N}  n^3\,K_2(n\,\lambda_0\,v^2\,\beta^2\,(2k+1)),
\end{eqnarray}
and we have introduced the dimensionless quantities $a=q\,A\,\beta$ and $\lambda_0 = \lambda/\beta $, the latter 
 representing the effective coupling at temperature $T$
($\lambda$ has units of $mass^{-1}$ since $[W]=3$).
The massless contribution (\ref{masacero}) is independent of $v$, therefore it does not need to be considered in looking for the extrema of the potential.

In order to study the  vacua of the theory in the general case  we look for solutions to $0=\frac{d V_{\rm eff}}{dv}$. This gives the equation

\bea
\label{eqextrema}
0= \frac{a^2\,v\,N\,(N+1)}{\beta^2}
+\frac{8\,\lambda_0^3\,\beta^2\,v^5}{\pi^2}\sum_{k=0}^{\infty}\,\frac{\cos(a\,(2k+1))}{(2k+1)}\,\sum_{n=1}^{N}  n^4\,K_1(n\,\lambda_0\,v^2\,\beta^2\,(2k+1))\ .
\eea

\noindent Let us first consider the finite-temperature properties of the standard fuzzy sphere described in \cite{Maldacena:2009mw}, that is, when the Wilson line is absent. When $a=0$, only the second term in (\ref{V}) remains. 
The way the effective potential (\ref{V}) behaves as a function of $v^2$ can be seen from the fact that 
the function $-z^2\,K_2(z)$ is a monotonically increasing function starting with $-2$ at $z=0$ and approaching $0$ asymptotically.
This implies that at $a=0$ the only extremum of the potential is the absolute minimum at $v=0$.
This can be verified directly by setting $a=0$ in (\ref{eqextrema}): then the first term in   (\ref{eqextrema}) vanishes and the second term is positive definite, which implies that
the only extremum of the potential is indeed at $v=0$.
Thus we conclude that the parameter $v$, a modulus at zero temperature, is driven towards 0 by finite temperature effects. 

\medskip 

In the following we show that, when the Wilson line is present ($a\neq 0$), in a certain range of the parameters the trivial solution $v=0$ will not be the absolute minimum of the potential. The non-trivial minimum arises in different intervals of $a$ and temperature, with $\lambda_0$ and $N$ above some minimum value.

\subsection{ Finite $N$ case}

We first consider finite (non-large) $N$ and look for solutions with $v\neq 0$ by suitably choosing the Wilson line $A$. Equation (\ref{eqextrema}) can be analytically studied in two opposite approximations depending on the argument of the Bessel function. 

\subsubsection*{$N\,\lambda_0\, v^2\,\beta^2\ll 1$} 

For small argument $z\ll 1$ the Bessel function behaves as $K_1(z)\sim 1/z +\mathcal{O}(1)$. 
This approximation can be used for the first few terms in the Bessel functions of (\ref{eqextrema}) when $N\,\lambda_0\, v^2\,\beta^2\ll 1$.
{}From some $k$ on, the argument of the Bessel function will not be small and this approximation cannot be used. However the $(2\,k+1)^2$ suppression for those terms (along with the fact that $z\,K_1(z)<1$) makes their contribution in fact negligible. Thus, for $N\,\lambda_0\, v^2\,\beta^2\ll 1$ eq. (\ref{eqextrema}) can be replaced by

\bea
0= a^2 +\frac{2\,N\,(N+1)\,\lambda_0^2\,v^2\,\beta^2}{\pi^2}\sum_{k=0}^{\infty} \frac{\cos(a\,(2k+1))}{(2k+1)^2}\ .
\eea

\noindent As discussed, the $k$-suppression allows us to consider the first few terms. In particular, keeping the first two terms one obtains

\be
v^2= -\frac{\pi^2\, a^2}{2\,N\,(N+1)\lambda_0^2\, \beta^2\,\left(\cos(a)+ {\cos(3a)\over 9}\right)}\ .
\ee

Recalling that $a= qA/T$, we see that $v$ is real in a certain range of temperature, so this  represents an extremum of the potential --indeed,
a maximum, as can be seen by computing $V_{\rm eff}''(v)$.

\subsubsection*{$\lambda_0\, v^2\,\beta^2\gg 1$}

Let us now consider the opposite limit, in which $\lambda_0\, v^2\,\beta^2\gg 1$. 
The Bessel functions can now be approximated by its asymptotic form, $K_1(z)\cong \sqrt{\pi\over 2z} \ e^{-z}$. Because of the
exponential suppression, we  keep only the leading term, \textit{i.e.} $k=0$, $n=1$. Thus

\bea
\label{ara}
0= a^2\,N\,(N+1)
+\frac{4\,\sqrt{2}\,\lambda_0^{5\over 2}\,\beta^3\,v^3}{\pi^{3\over 2}}\,\cos(a)\,e^{-\lambda_0\,v^2\,\beta^2}\ .
\eea
 With no loss of generality  we have assumed $v>0$, as the potential only depends on $v^2$. Then (\ref{ara}) has solutions provided $\cos(a)<0$, which, within this approximation, implies  

\be
\label{intervals}
(2\,\ell+{1\over2})\,\pi  < a < (2\,\ell+{3\over2})\,\pi  \qquad {\it i.e.} \qquad 
 \frac{q\,A}{(2\,\ell+{1\over 2})\,\pi } > T > \frac{q\,A}{(2\,\ell+{3\over 2})\,\pi } \ ,\qquad \ell=0,1,2,...
\ee
Solutions exist only in a finite number of these intervals.
The equation defining the extrema can have two roots, one double root or none.
 An analysis 
shows that the equation has solutions provided $a$ is less than some maximum value $\sim \sqrt{\lambda_0}$. 
  Therefore, the intervals in (\ref{intervals}) which give rise to solutions are $\ell=0,1,...,\ell_{\rm max}$ with
 $\ell_{\rm max}\sim \sqrt{\lambda_0}\ $.

\bigskip 

We can now combine the above information to understand the behavior of the potential. In any of the allowed intervals (\ref{intervals}), for a sufficiently large $\lambda_0$, there are two extrema besides the trivial one $v=0$, a maximum and a minimum, in fact captured respectively by the two regimes above: the one with lower $v$ represents a maximum and the one with higher $v$ represents a relative minimum of $V_{\rm eff}$, which becomes an absolute minimum if $\lambda_0$ is sufficiently large.

\begin{figure}[h!]
\centering
\subfigure{\includegraphics[width=6.5 cm]{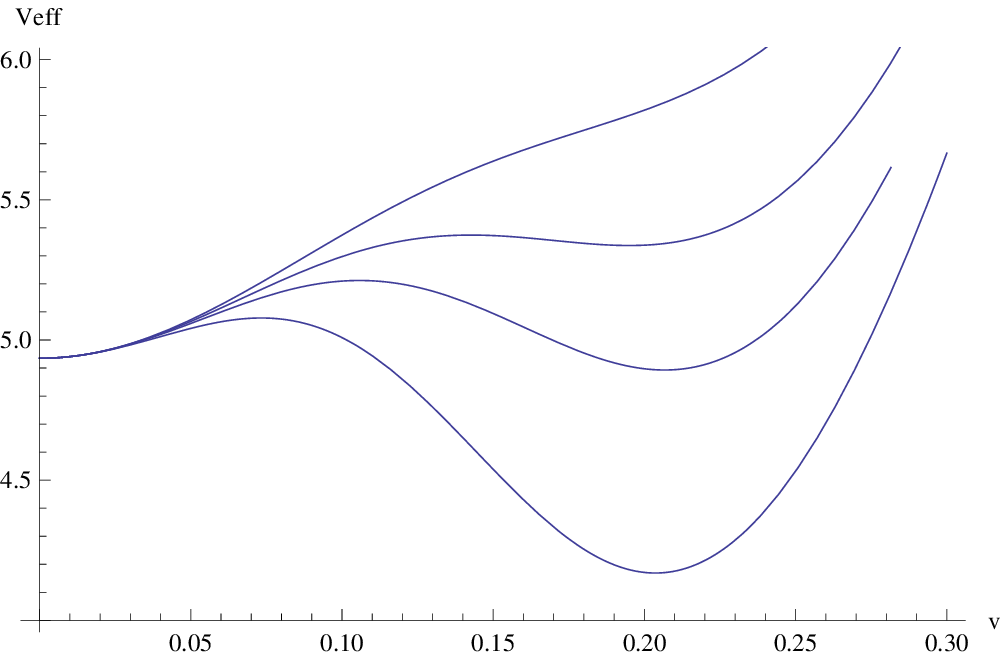}}
\ \ \ \ 
\subfigure{\includegraphics[width=6.5cm]{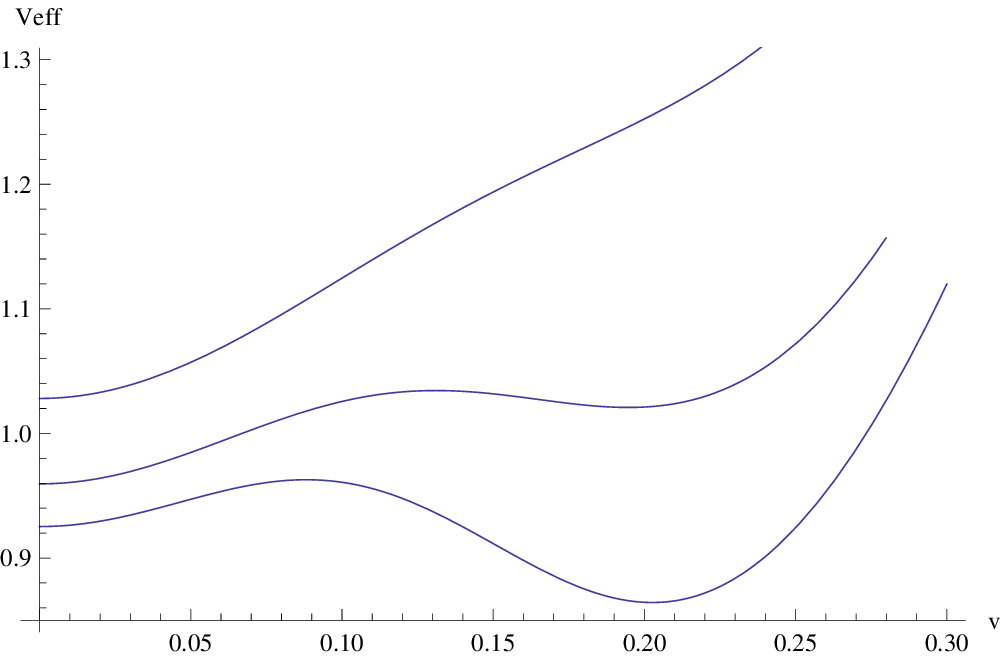}}
\caption{ \footnotesize (a) $V_{\rm eff}$ vs. $v$ for $N=2$, $a=\pi$, $\beta=1$ and $\lambda_0=20,25,30,40$. The potential develops a maximum and a minimum
for $\lambda_0>23.7$. For $\lambda_0>29.5$ the non-trivial minimum becomes also the absolute minimum.
(b)  $V_{\rm eff}$ vs. $v$ for $\lambda_0=10$, $a=\pi $, $\beta=1$ and $N=4,6,8$.
\label{effpotential}}
\end{figure}

The exact potential is shown in fig. \ref{effpotential}. One can see that non-trivial maximum and minimum appear for $\lambda_0$ or $N$ above some critical value (provided $a$ is in a certain interval).
We should note that from fig. \ref{effpotential}.(b) it follows that the non-trivial minimum becomes deeper as $N$ is increased fixing $\lambda_0$.

\begin{figure}[h!]
\centering
\subfigure{\includegraphics[width=6. cm]{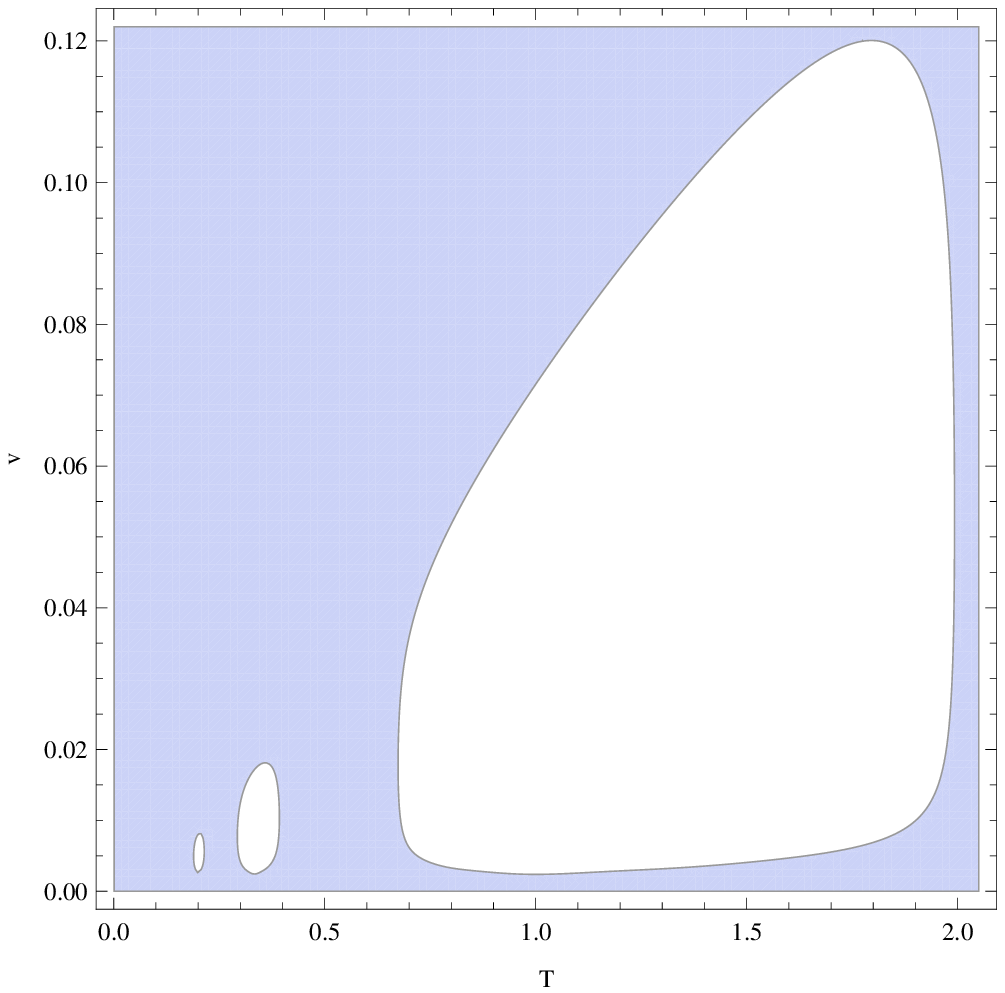}}
\ \ \ 
\subfigure{\includegraphics[width=6.cm]{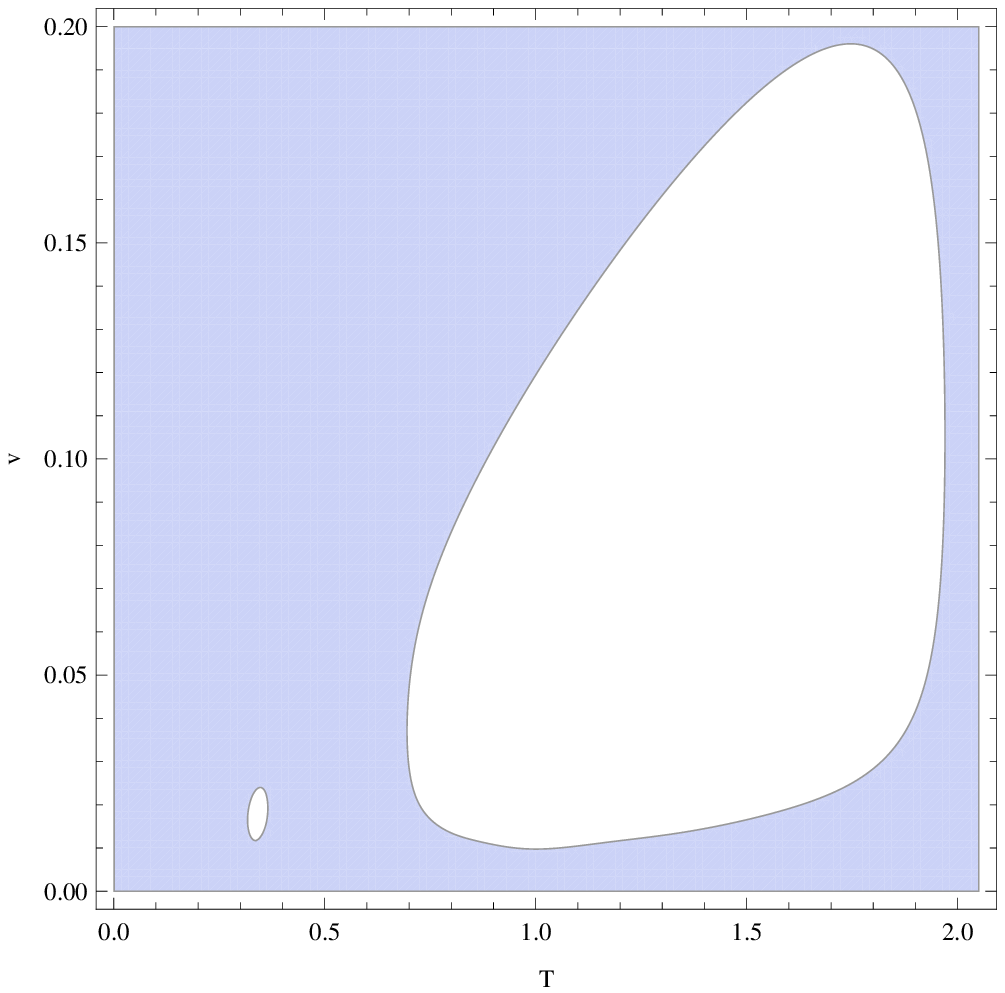}}
\caption{ \footnotesize (a) The solutions $v(T)$ of $\frac{dV_{\rm eff}}{dv}=0$ for $q\,A= \pi$, $N=2$, $\lambda_0=1100$. The lower
parts of the boundary of each bubble describe maxima and the upper parts describe minima of $V_{\rm eff}$.
(b) Same for $\lambda_0=270$. The third (smallest) bubble has disappeared and the second bubble is about to disappear.
The largest bubble disappears for $\lambda_0< 20$.
\label{critical}}
\end{figure}

The values $v(T)$ representing the extrema of the potential are  shown in fig. \ref{critical} for $q\, A=\pi$. The lower part of the boundaries of the bubbles represent the maxima, whereas the upper part of the boundaries represent the minima. In general, for sufficiently large $\lambda_0$ or $N$, there is a large bubble and a number of smaller ones. Each bubble lies  within a window of $T$ which at large $\lambda_0$ corresponds to one of the
intervals in (\ref{intervals}). The number of bubbles is reduced if  $\lambda_0$ is reduced. Eventually also the largest bubble in
fig. \ref{critical}.(a) becomes small and disappears as $\lambda_0 $ gets smaller than some critical value, which depends on $q\,A$. Figs. \ref{critical}.(a) and \ref{critical}.(b) also exhibit the fact that the minimal value of $\lambda_0$ below which there is no extrema depends on the temperature.
For example, this can be seen  gradually decreasing
$\lambda_0$ and considering the effect on the two extrema $v(T)$ in fig.  \ref{critical}. The bubbles become narrower, the minimum and maximum at a given $T$ approach each other until they merge and eventually disappear at some critical
$\lambda_0(T)$.

\medskip

\begin{figure}[h!]
\centering
\subfigure{\includegraphics[width=6. cm]{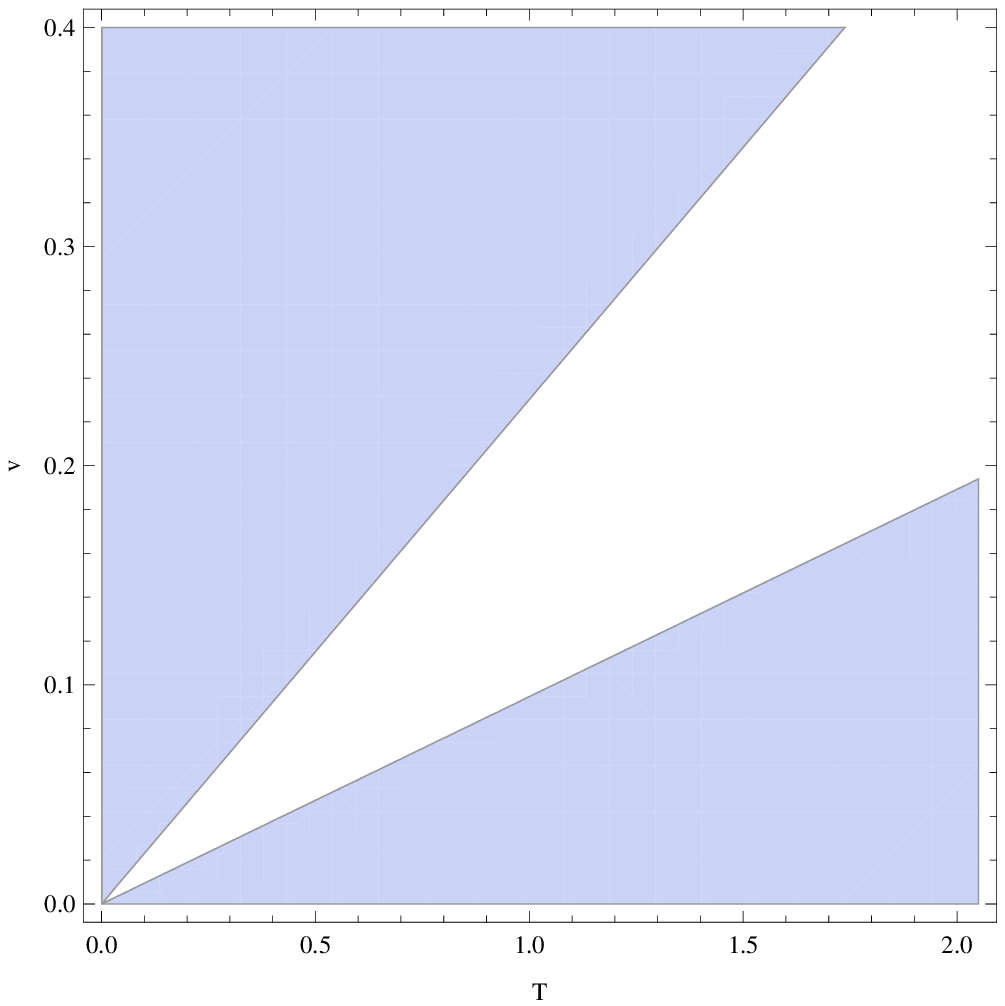}}
\ \ \ 
\subfigure{\includegraphics[width=6.cm]{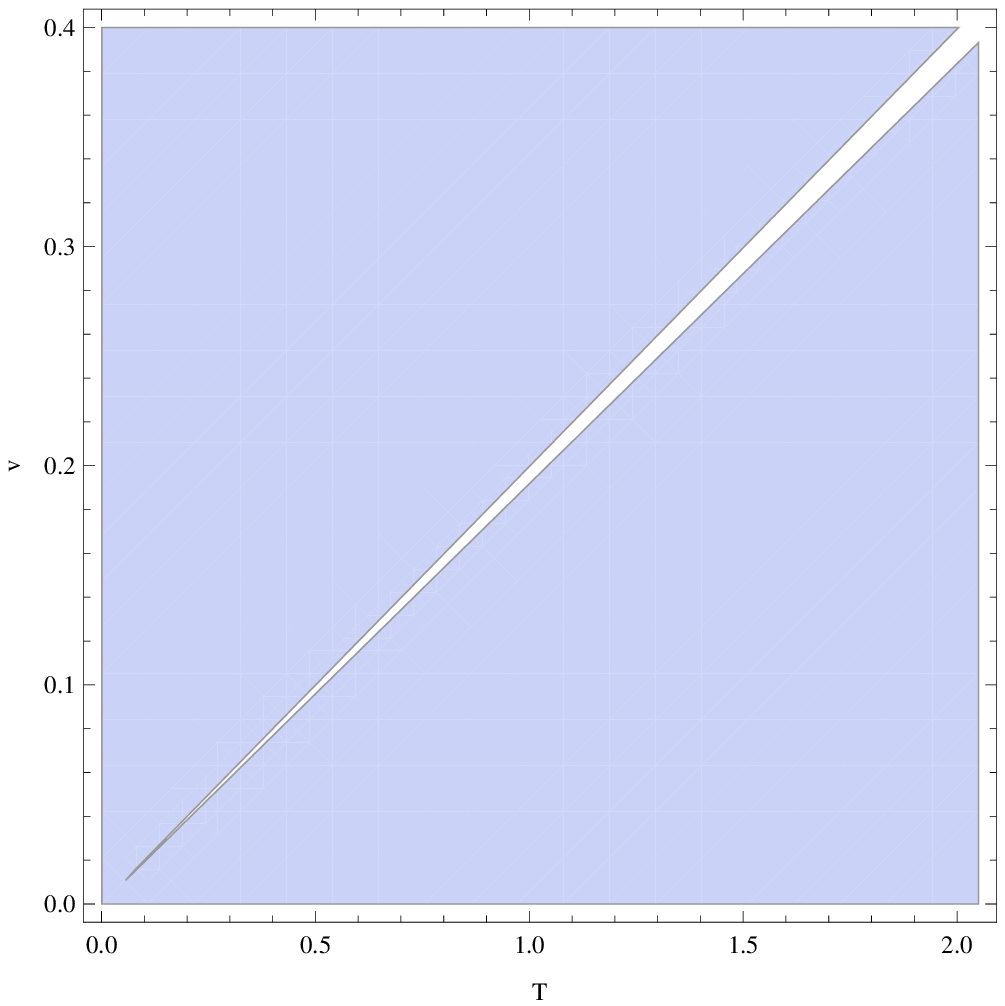}}
\caption{ \footnotesize  The solutions $v(T)$ of $\frac{dV_{\rm eff}}{dv}=0$, this time shown at fixed $a$. The upper (lower) line represents the minima (maxima) at 
different temperatures.
a) For $a=0.8\,\pi$, $\lambda_0=30$ at $N=2$; b)  For $a=0.8\,\pi$, $\lambda_0=19.2$ at $N=2$. For lower values of $\lambda_0$ the only extremum is the trivial minimum at $v=0$.
 \label{fixeda}}
\end{figure}

So far we have considered fixing the value of the Wilson line parameter $A$. As the effective potential comes in terms of the dimensionless quantity $a=q\,A\,\beta$, it is instructive  to consider the dynamics at fixed $a$, \textit{i.e.} scaling the Wilson line with the temperature so that $a$ is kept fixed. The limiting lines of the wedge in figure \ref{fixeda} show the maximum (lower line) and the minimum (upper line) of the effective potential at various temperatures fixing $a$. These plots have been done at fixed $\lambda_0$. As shown in fig. \ref{fixeda}, as $\lambda_0$ is decreased the wedge closes. Indeed, below $\lambda_0\sim 20$ the wedge has completely closed, that is, as anticipated we loose the non-trivial extrema if we consider a too weak interaction. This is expected since the present effect arises from a balance between a tree-level and a one-loop contribution.

\begin{figure}[h!]
\centering
\subfigure{\includegraphics[width=6. cm]{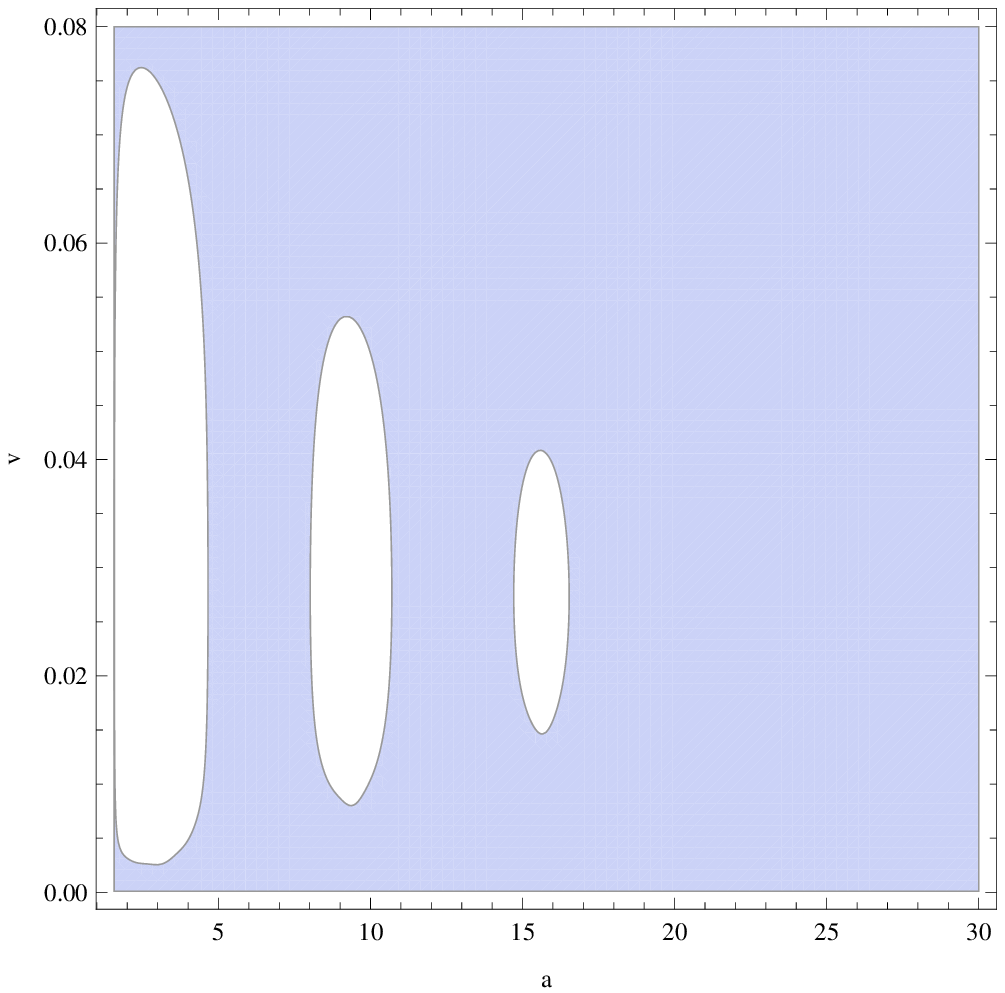}}
\ \ \ 
\subfigure{\includegraphics[width=6.cm]{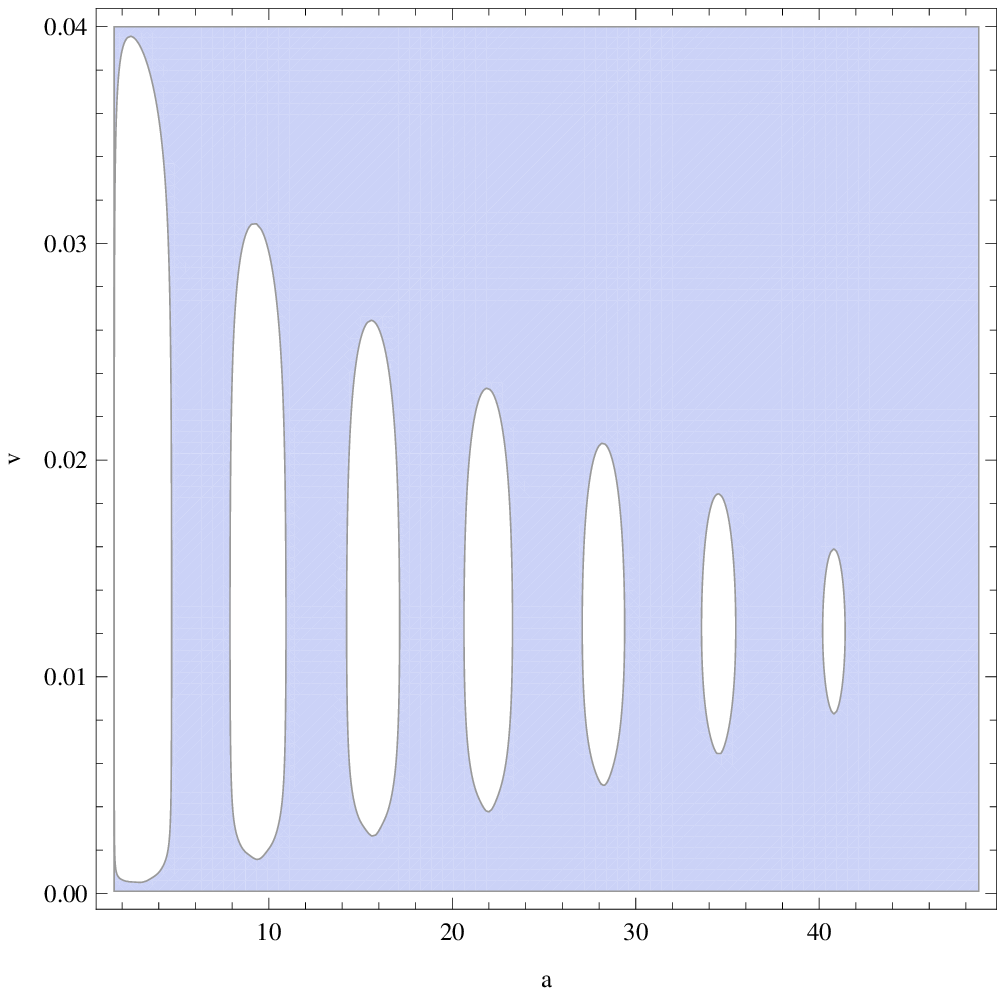}}
\caption{ \footnotesize  The solutions $v(a)$ of $\frac{dV_{\rm eff}}{dv}=0$, now at fixed $T=1$, $N=2$. 
a) $\lambda_0=1000$; b) $\lambda_0=5000$.
 \label{fixedT}}
\end{figure}

One can also fix $\lambda_0$ and consider the behavior of the wedge as $a$ is changed. For example, for $\lambda_0=30$, the wedge exists for $a\in[0.596\,\pi,\, 1.095\,\pi]$ (and it is widest for $a\cong 0.8\, \pi$), thus in reasonably good agreement with (\ref{intervals}). 
The agreement becomes better for
larger values of $\lambda_0$. Figs. \ref{fixedT}.(a), \ref{fixedT}.(b) show $v(a)$ at fixed $T=1$ and $\lambda_0=1000,\ 5000$.
One can see the presence of a finite number of intervals which increases as $\lambda_0$ increases.
While for simplicity of the exposition we used the special point $a=\pi$ to illustrate the mechanism, it is now clear that it applies for a range of values of $a$. 
The effect is in fact stronger  for 
 $a\cong 0.8\, \pi$, where the non-trivial minimum is deepest (in particular, in the interval 
$19.2<\lambda_0< 23.7$, only the trivial vacuum $v=0$ exists for $a=\pi$, while there are still non-trivial extrema at $a\cong 0.8\, \pi$).

\subsection{ Large $N$ case}

{}For large $N$ we can approximate the $n$-sum in terms of an integral by introducing $x\equiv n/N$ so that

\begin{equation}
\sum_{n=1}^N n^3\,K_2(n\,\lambda_0\, v^2\,\beta^2\, (2k+1))\sim N^4\int_0^1 dx \, x^3\, K_2(x \, y)\ ,
\end{equation}
with $y\equiv g\,v^2\beta^2\,(2k+1) $. Computing the integral we find

\begin{equation}
\int_0^1 dx\,x^3\,K_2(x \ y)=\frac{1}{y^4}\, F(y)\ ,\qquad F(y)=8-4\,y^2\,K_0(y) - y\,(8+y^2)\,K_1(y)\ .
\end{equation}
Then, the effective potential becomes 

\begin{equation}
\label{tHooft}
N^{-2}\,\beta^2\,V_{\rm eff}= a^2v^2 -\frac{8}{\pi^2\,g^2\,v^4\beta^6} \sum_{k=0}^\infty \frac{\cos(a(2k+1))}{(2k+1)^6}\,F(g\,v^2\beta^2 (2k+1))\ .
\end{equation}
Thus, in the large $N$ limit with fixed $g=\lambda_0\,N$, all terms in $V_{\rm eff}$ are homogeneous in $N$. 
In the one-loop approximation, the existence of non-trivial minimum is therefore independent of $N$, it only depends on the 't Hooft coupling $g$.
More generally,  since the Lagrangian (\ref{uno}) admits a standard large $N$ expansion, one expects that the location of the
non-trivial minimum should be independent of $N$ to all loop 
orders in the planar approximation.

The existence of non-trivial maxima and minima of the potential  now becomes sensitive to the value of $g$.
For fixed $a$ and $T$, one can see that there is a minimum value of $g$ above which $V_{\rm eff}$ has maximum and a minimum,
in a similar way as in the $N=\mathcal{O}(1)$ case. We omit the figures, as $V_{\rm eff}(v)$ has a qualitative shape which is essentially 
the same as in fig. \ref{effpotential}a, now plotted for different values of $g$. The counterparts of figs. \ref{critical},  \ref{fixeda} and  \ref{fixedT} are also similar, with $g$ playing the role of $\lambda_0$.

\subsection{Comments on the regime of validity of the approximation}

Let us now pause to consider the regime of validity of our approximation. The coupling constant appearing in the (component) lagrangian is $\lambda^2$. As $\lambda$ is a dimensionful parameter, the perturbative expansion will be  in powers of a dimensionless parameter $(\lambda\, E)^2$, being $E$ the energy of the process. In the present  case, since the system is at finite temperature, we can estimate $E$ as the average energy per particle in the heat bath, that is $E=\frac{T}{2}$.  
Furthermore, as it is well-known (see \textit{e.g.} section 6.2.3 of \cite{IZ}), $L$ loop integrals yield further $(4\pi)^{-2\,L}$ suppressing factors. There is also an extra factor of $2^{-3/2}$ coming from canonically normalizing the fields. Taking these factors into account, 
the perturbative expansion should be in powers of the coupling
\begin{equation}
\kappa=\frac{\lambda_0^2}{2^{15/2}\,\pi^2}\ ,\qquad \lambda_0=\lambda\, T\ .
\end{equation}
Using this one finds a window where our one-loop approximation can in principle be justified ($\kappa<1$) and at the same time can lead to non-trivial effects
We should note however that the the coupling does not go deep into the perturbative regime, as our effect comes from balance between one loop and classical effects (and therefore they must be competing). 
Nonetheless, we expect that one effect of higher loops would be  to increase the mass of the fluctuations, in this way weighting  the quantum contribution  more. As a result, the threshold value of $\lambda_0 $ above which a non-trivial minimum of the effective potential exists would be smaller.

\section{Conclusions}\label{conclusions}

At zero temperature,  the radius of the fuzzy sphere solution found in \cite{Maldacena:2009mw} is a modulus which stands for a scalar component in the baryonic current multiplet. At finite  temperature,  we found that one-loop effects drive this parameter towards zero value, {\it i.e.} the fuzzy 2-sphere is shrunk to vanishing radius.
Although our calculation is perturbative, there is a
 related effect at strong coupling: as shown in \cite{Buchel:2001gw, Gubser:2001ri} by the analysis of the gravity dual, at temperatures well above the strong coupling scale both chiral and baryonic symmetries are restored, a condition which is satisfied by our weakly coupled analysis.

By deforming the theory adding a background (\textit{i.e.} non-dynamical) gauge field for a global symmetry --~such as the $U(1)_R$ or the $U(1)_B$~--, we proposed a mechanism for stabilization of the fuzzy sphere using similar ingredients as in the Hosotani mechanism \cite{Hosotani:1983xw}. The key observation is that there is a window in the Wilson loop parameter where the one-loop potential flips sign, rendering the $v=0$ vacuum metastable and leading to an absolute minimum at some $v\neq 0$.

Other interesting applications of the present  mechanism are in inflationary cosmology. Thinking of $v$ as an inflaton, it would roll down the potential towards the non-trivial minimum and back to $v=0$ if the temperature is lowered. In fact, this has some similarities with the scenarios discussed in  \cite{DeWolfe:2004qx,Ashoorioon:2009wa}.
It is interesting to note that the same mechanism for generation of non-trivial vacua by Wilson lines arises in more general settings, which have
no connection with KS or fuzzy spheres. One could simply consider Einstein gravity coupled to a complex scalar and a fermion described e.g. by a Wess-Zumino Lagrangian with a $U(1)$ symmetry and some superpotential that provides equal masses for them once the scalar takes a vacuum expectation value. Turning on a temporal Wilson line at finite temperature, would then lead to a picture like in figure \ref{critical} with multiple intervals of temperature. The existence of these intervals means that, if the temperature is lowered, the universe undergoes different inflationary periods for each window of temperature. Each successive period has an inflaton field with lower value (determined by the upper part of the boundary of the bubbles of figure \ref{critical} and a cosmological constant parameter (determined by the value of the effective potential) that gets substantially smaller at each successive inflationary period. We leave this very interesting problem open for future research.

Finally, it would be interesting to study whether this mechanism survives to strong coupling. The vanshing Wilson line case has been studied in the past at strong coupling from a gravitational perspective (see \textit{e.g.} \cite{Buchel:2001gw, Gubser:2001ri} as well as the recent work \cite{Caceres:2011zn}). As for the $A\ne 0$ case, while we have exhibited the effect in the weak coupling approximation (as well as  in the large $N$ weak 't Hooft coupling approximation), this effect seems to become stronger for larger couplings. To investigate the strong coupling regime, a holographic approach is best suited. As the gauge fields in $AdS_5$ dual to the baryonic and R-symmetry currents are known, it might be possible to study this phenomenon from the gravitational point of view. In fact a very interesting discussion on the role of background gauge fields for  global symmetries and its relation to the gauge/gravity duality has recently appeared in \cite{Maeda:2010br}, where a prescription is discussed for the gravitational realization of  deformations such as the one of the present work. 
For example, for the $U(1)_R$, in the present case one can attempt a direct construction:
an ansatz for the  gravitational dual background can be constructed in terms
of Melvin twists applied to the KS finite temperature background, mixing the time circle with the $U(1)$ circle in the $T^{1,1}$ base. Such twists may generate Wilson lines for a $U(1)$ background gauge field of
the boundary theory and typically  produce  twisted boundary conditions on fermions and bosons on one circle \cite{Russo:1995ik}.\footnote{A charged field has the form $\Psi =e^{iq Y}\psi(x)$, where $Y$ parametrizes the $U(1)$ circle. A redefinition $Y'=Y+A\tau$ in the finite temperature KS metric, where $\tau $ is the Euclidean time, then produces the desired twisted boundary conditions, in this case on the $\tau$-circle.}
We leave this open for future work.

\section*{Acknowledgments}

D.R-G. wishes to thank the University of Oviedo for warm hospitality during the initial stages of this work. D.R-G is supported by the Israel Science Foundation through grant 392/09. He also acknowledges support from the Spanish Ministry of Science through the research grant FPA2009-07122 and Spanish Consolider-Ingenio 2010 Programme CPAN (CSD2007-00042). J.R.~acknowledges support by MCYT Research
Grant No.  FPA 2010-20807-C02-01 and project 2009SGR502.


\begin{thebibliography}{99}

\bibitem{Klebanov:1998hh}
  I.~R.~Klebanov, E.~Witten,
  ``Superconformal field theory on three-branes at a Calabi-Yau singularity,''
  Nucl.\ Phys.\  {\bf B536 } (1998)  199-218.
  [hep-th/9807080].
  
\bibitem{Klebanov:2000hb}
  I.~R.~Klebanov, M.~J.~Strassler,
  ``Supergravity and a confining gauge theory: Duality cascades and chi SB resolution of naked singularities,''
  JHEP {\bf 0008}, 052 (2000).
  [hep-th/0007191].
  
\bibitem{Butti:2004pk}
  A.~Butti, M.~Grana, R.~Minasian, M.~Petrini, A.~Zaffaroni,
  ``The Baryonic branch of Klebanov-Strassler solution: A supersymmetric family of SU(3) structure backgrounds,''
  JHEP {\bf 0503}, 069 (2005).
  [hep-th/0412187].
  
\bibitem{Dymarsky:2005xt}
  A.~Dymarsky, I.~R.~Klebanov, N.~Seiberg,
  ``On the moduli space of the cascading SU(M+p) x SU(p) gauge theory,''
  JHEP {\bf 0601}, 155 (2006).
  [hep-th/0511254].
  
\bibitem{Maldacena:2009mw}
  J.~Maldacena, D.~Martelli,
  ``The Unwarped, resolved, deformed conifold: Fivebranes and the baryonic branch of the Klebanov-Strassler theory,''
  JHEP {\bf 1001}, 104 (2010).
  [arXiv:0906.0591 [hep-th]].
  
\bibitem{Hosotani:1983xw}
  Y.~Hosotani,
  ``Dynamical Mass Generation by Compact Extra Dimensions,''
  Phys.\ Lett.\  B {\bf 126} (1983) 309.
  
\bibitem{Scherk:1978ta}
  J.~Scherk, J.~H.~Schwarz,
  ``Spontaneous Breaking of Supersymmetry Through Dimensional Reduction,''
  Phys.\ Lett.\  {\bf B82}, 60 (1979).
  
\bibitem{Scherk:1979zr}
  J.~Scherk, J.~H.~Schwarz,
  ``How to Get Masses from Extra Dimensions,''
  Nucl.\ Phys.\  {\bf B153}, 61-88 (1979).
  
\bibitem{DeWolfe:2004qx}
  O.~DeWolfe, S.~Kachru, H.~L.~Verlinde,
  ``The Giant inflaton,''
  JHEP {\bf 0405}, 017 (2004).
  [hep-th/0403123].
   
\bibitem{Ashoorioon:2009wa}
  A.~Ashoorioon, H.~Firouzjahi, M.~M.~Sheikh-Jabbari,
  ``M-flation: Inflation From Matrix Valued Scalar Fields,''
  JCAP {\bf 0906}, 018 (2009).
  [arXiv:0903.1481 [hep-th]].
  
\bibitem{Aharony:1997bx}
  O.~Aharony, A.~Hanany, K.~A.~Intriligator, N.~Seiberg, M.~J.~Strassler,
  ``Aspects of N=2 supersymmetric gauge theories in three-dimensions,''
  Nucl.\ Phys.\  {\bf B499}, 67-99 (1997).
  [hep-th/9703110].
  
\bibitem{Festuccia:2011ws}
  G.~Festuccia, N.~Seiberg,
  ``Rigid Supersymmetric Theories in Curved Superspace,'' 
  [arXiv:1105.0689 [hep-th]].
  
\bibitem{Maeda:2010br}
  K.~Maeda, M.~Natsuume, T.~Okamura,
  ``On two pieces of folklore in the AdS/CFT duality,''
  Phys.\ Rev.\  {\bf D82 } (2010)  046002.
  [arXiv:1005.2431 [hep-th]].
  
\bibitem{Dolan:1973qd}
  L.~A.~Dolan, R.~Jackiw,
  ``Symmetry Behavior at Finite Temperature,''
  Phys.\ Rev.\  {\bf D9}, 3320-3341 (1974).
  
    \bibitem{IZ}
  C.Itzykson and B.Zuber, \textit{Quantum Field Theory}, McGraw-Hill, 1980.
  
\bibitem{Buchel:2001gw}
  A.~Buchel, C.~P.~Herzog, I.~R.~Klebanov, L.~A.~Pando Zayas, A.~A.~Tseytlin,
  ``Nonextremal gravity duals for fractional D-3 branes on the conifold,''
  JHEP {\bf 0104}, 033 (2001).
  [hep-th/0102105].
  
\bibitem{Gubser:2001ri}
  S.~S.~Gubser, C.~P.~Herzog, I.~R.~Klebanov, A.~A.~Tseytlin,
  ``Restoration of chiral symmetry: A Supergravity perspective,''
  JHEP {\bf 0105 } (2001)  028.
  [hep-th/0102172].
  
  
\bibitem{Caceres:2011zn}
  E.~Caceres, C.~Nunez, L.~A.~Pando-Zayas,
  ``Heating up the Baryonic Branch with U-duality: A Unified picture of conifold black holes,''
  JHEP {\bf 1103 } (2011)  054.
  [arXiv:1101.4123 [hep-th]].
  
\bibitem{Russo:1995ik}
  J.~G.~Russo and A.~A.~Tseytlin,
  ``Magnetic flux tube models in superstring theory,''
  Nucl.\ Phys.\  B {\bf 461}, 131 (1996)
  [arXiv:hep-th/9508068].


\end{thebibliography}
\end{document}